\documentclass[aip,cha,amsmath,amssymb,reprint,author-year,author-numerical]{revtex4-1}
\usepackage{epsfig}
\usepackage{color}
\usepackage{graphicx}
\usepackage{dcolumn}
\usepackage{bm}
\input epsf

\usepackage[normalem]{ulem} 

\graphicspath{{figures/}}

\begin{document}


\title{Tiered synchronization in coupled oscillator populations with interaction delays and higher-order interactions}

\author{Per Sebastian Skardal}
\email{persebastian.skardal@trincoll.edu}
\affiliation{Department of Mathematics, Trinity College, Hartford, CT 06106, USA}

\author{Can Xu}
\affiliation{Institute of Systems Science and College of Information Science and Engineering, Huaqiao University, Xiamen 361021, China}

\begin{abstract}
We study synchronization in large populations of coupled phase oscillators with time-delays, higher order interactions. With each of these effects individually giving rise to bistabiltiy between incoherence and synchronization via a subcriticality at the onset of synchronization and the development of a saddle node, we find that their combination yields another mechanism behind bistability, where supercriticality at onset may be maintained and instead the formation of two saddle nodes creates tiered synchronization, i.e., bistability between a weakly synchronized state and a strongly synchronized state. We demonstrate these findings by first deriving the low dimensional dynamics of the system and examining the system bifurcations using a stability and steady-state analysis.
\end{abstract}


\maketitle

\begin{quotation}
Collective oscillations in large populations of coupled dynamical units play a critical role in applications in mathematics, physics, engineering, and biology~\cite{Strogatz2003,Pikovsky2003}. Examples where robust synchronization plays a role in system function includes Josephson junction arrays~\cite{Wiesenfeld1996PRL}, cardiac tissue~\cite{Glass1988}, circadian clocks~\cite{Yamaguchi2003Science}, and the power grid~\cite{Rohden2012PRL}. Two properties that have been shown add richness to a systems' macroscopic dynamics are interaction delays~\cite{Lee2009PRL} and higher-order interactions~\cite{Skardal2020CommPhys}, both of which induce bistability and abrupt synchronization transitions. Here we examine the dynamics of coupled oscillator populations with both interactions delays and higher order interactions present. In addition to the development of a subcriticality, as is the the typical source of bistabiltiy in previous work, we find that the combination of these two effects promotes bistability via a double saddle-node bifurcation as the onset of synchronization remains supercritical, thereby giving rise to states of tiered synchronization where both weakly and strongly synchronized states coexist.
\end{quotation}

\section{Introduction}\label{sec1}

In the interdisciplinary study of collective behavior and synchronization, the Kuramoto model is of pivotal importance due to it analytical tractability and versatility for modeling a wide range of behaviors~\cite{Kuramoto1984}. This versatility comes in part from its ability to incorporate a wide range of properties critical for different physical and biological systems. One such property is the presence of interaction delays~\cite{Kim1997PRL,Yeung1999PRL,Choi2000PRE}, which give rise to rich dynamics and multistability. Another property that has attracted more attention in recent years is the presence of higher-order interactions~\cite{Horak2009,Otter2017EPJDS,Salnikov2019EJP,Carletti2020JPhys,Battiston2020PhysRep}, most notably motivated by applications neuroscience~\cite{Petri2014Interface,Giusti2016JCN,Reimann2017,Sizemore2018JCN} and physics~\cite{Ashwin2016PhysD,Leon2019PRE}. In fact, the effect of higher-order interactions have already been investigated in both synchronization~\cite{Tanaka2011PRL,Komarov2015PRE,Bick2016Chaos,Skardal2019PRL,Xu2020PRR,Skardal2020JPhys,Millan2020PRL,Mulas2020PRE,Lucas2020PRR,Xu2021PRR,Skardal2021PRR,Wang2021PRE} and other kinds of collective behavior~\cite{Schaub2019,Ziegler2022Chaos,Iacopini2019NatComms,Matamalas2019}.

Our understanding of the macroscopic dynamics of coupled oscillator populations with interactions delays and with higher order interactions has been further developed by applying the dimensionality reduction of Ott and Antonsen~\cite{Ott2008Chaos,Ott2009Chaos}. In the case of heterogeneous interaction delays, as the characteristic time delay and mean natural frequency are increased the onset of synchronization is likewise made larger, eventually transitioning from a supercritical bifurcation to a subcritical one as the curve of steady-state solutions folds over itself and a saddle-node bifurcation is born~\cite{Lee2009PRL,Lee2011Chaos,Laing2011PhysD,Skardal2014PhysD,Skardal2018IJBC}. On the other hand, in systems with higher order interactions, the higher-order interactions themselves do not alter the location of the onset of synchronization or the stability of the incoherent state, but promote synchronization via nonlinear terms~\cite{Skardal2020CommPhys}. Increasing the higher order coupling strength eventually causes the curve of steady-state solutions to similarly fold over itself as a saddle-node bifurcation is born. Thus, in both cases bistability emerges between the incoherent and synchronized states.

In this paper we study populations of coupled phase oscillators with both interaction delays and higher order interactions. The dynamics incorporate dyadic, triadic, and tetradic interactions, with heterogeneous time delays between each pair of oscillators. After applying the dimensionality reduction of Ott and Antonsen and analyzing the reduced macroscopic dynamics, we find that bistabilty remains a key feature, but the combination of interaction delays and higher order interactions promotes an additional mechanism than that described above. Specifically, in addition to bistability between the incoherent state and the synchronized state owing to a subcritical pitchfork and a saddle node, we also observe a pair of saddle nodes that leaves the onset of synchronization supercritical, leading to tiered synchronization, i.e., bistability between a weakly synchronized state and a strongly synchronized state.

The remainder of this paper is organized as follows. In Sec.~\ref{sec2} we present the governing equations and the dimensionality reduction using the Ott-Antonsen ansatz. In Sec.~\ref{sec3} we present an analysis of the steady-state dynamics. In Sec.~\ref{sec4} we analyze the incoherent state. In Sec.~\ref{sec5} we consider the special case of eliminating triadic coupling, leading to further analytical results and a sketch of an illustrative bifurcation diagram. In Sec.~\ref{sec6} we conclude with a discussion of our results.

\section{Governing Equations and Model Reduction}\label{sec2}

In this work we consider populations of coupled phase oscillators with both interaction delays and higher order interactions. We consider heterogeneous time delays between oscillators~\cite{Lee2009PRL} with dyadic, triadic, and tetradic interactions~\cite{Skardal2020CommPhys}, yielding
\begin{widetext}
\begin{align}
\dot{\theta}_i=\omega_i&+\frac{K_1}{N}\sum_{j=1}^N\sin[\theta_j(t-\tau_{ij})-\theta_i(t)]+\frac{K_2}{N^2}\sum_{j=1}^N\sum_{l=1}^N\sin[2\theta_j(t-\tau_{ij})-\theta_l(t-\tau_{il})-\theta_i(t)]\nonumber\\
&+\frac{K_3}{N^3}\sum_{j=1}^N\sum_{l=1}^N\sum_{m=1}^N\sin[\theta_j(t-\tau_{ij})+\theta_l(t-\tau_{il})-\theta_m(t-\tau_{im})-\theta_i(t)],\label{eq:01}
\end{align}
\end{widetext}
where $\theta_i$ and $\omega_i$ are the phase and natural frequency of oscillator $i$, $K_1$, $K_2$, and $K_3$ are the respective 1-, 2-, and 3-simplex coupling strengths, and $\tau_{ij}$ is the interaction delay between oscillators $i$ and $j$ felt by oscillator $i$. In general we assume that natural frequencies and time delays are drawn from their respective distributions $g(\omega)$ and $h(\tau)$. While the degree of synchronization is measured by the magnitude $r$ or the the classical instantaneous Kuramoto order parameter, given by
\begin{align}
z=re^{i\psi}=\frac{1}{N}\sum_{j=1}^Ne^{i\theta_j(t)},\label{eq:02}
\end{align}
we also define the so-called Daido order parameter
\begin{align}
z^{(1)}=r^{(1)}e^{i\psi^{(2)}}=\frac{1}{N}\sum_{j=1}^Ne^{2i\theta_j(t)},\label{eq:02a}
\end{align}
as well as two different varieties of oscillator-specific time-delayed order parameters, given by
\begin{align}
w_i^{(1)}&=\rho_i^{(1)}e^{i\phi_i^{(1)}}=\frac{1}{N}\sum_{j=1}^Ne^{i\theta_j(t-\tau_{ij})},~\text{and}\label{eq:03}\\
w_i^{(2)}&=\rho_i^{(2)}e^{i\phi_i^{(2)}}=\frac{1}{N}\sum_{j=1}^Ne^{2i\theta_j(t-\tau_{ij})}.\label{eq:04}
\end{align}
Using Eqs.~(\ref{eq:03}) and (\ref{eq:04}) we may rewrite Eq.~(\ref{eq:01}) as
\begin{align}
\dot{\theta}_i=\omega_i&+\frac{K_1}{2i}\left(w_i^{(1)}e^{-i\theta_i(t)}-w_i^{(1)*}e^{i\theta_i(t)}\right)\nonumber\\
&+\frac{K_2}{2i}\left(w_i^{(2)}w_i^{(1)*}e^{-i\theta_i(t)}-w_i^{(2)*}w_i^{(1)}e^{i\theta_i(t)}\right)\nonumber\\
&+\frac{K_3}{2i}\left((w_i^{(1)})^2w_i^{(1)*}e^{-i\theta_i(t)}-(w_i^{(1)*})^2w_i^{(1)}e^{i\theta_i(t)}\right),\label{eq:05}
\end{align}
where $*$ denotes complex conjugate.

Next we seek to derive a closed-form system governing the dynamics of the order parameter, for which purpose we consider the continuum limit of infinitely-many oscillators, $N\to\infty$. Note that in this limit we may express the order parameter as the integral
\begin{align}
z(t)=\iint f(\omega,\theta,t)e^{i\theta(t)}d\theta d\omega,\label{eq:06}
\end{align}
where $f(\omega,\theta,t)$ is the density function that describes the fraction $f(\omega,\theta,t)d\theta d\omega$ of oscillators with phase and frequency, respectively, in $[\theta,\theta+d\theta)$ and $[\omega,\omega+d\omega)$ at time $t$. Moreover, the time-delayed order parameters may be written
\begin{align}
w_i^{(1)}(t)&=\iiint f(\omega,\theta,t-\tau)e^{i\theta(t-\tau)}h(\tau)d\theta d\omega d\tau,\label{eq:07}\\
w_i^{(2)}(t)&=\iiint f(\omega,\theta,t-\tau)e^{2i\theta(t-\tau)}h(\tau)d\theta d\omega d\tau.\label{eq:08}
\end{align}
Importantly, the form of Eqs.~(\ref{eq:07}) and (\ref{eq:08}), which may be rewritten $w_i^{(1,2)}(t)=\int z^{(1,2)}(t-\tau)h(\tau)d\tau$, (where $z^{(1)}(t)=z(t)$) reveals that in the continuum limit the variation between the time delayed order parameters vanish across different oscillators, so we may drop the subscripts, i.e., $w_i^{(1)}=w^{(1)}$ and $w_i^{(2)}=w^{(2)}$ for all $i$. I.e., in the continuum limit each time delayed order parameter $w_i(t)$ is defined by the same mean field via $z(t)$. With this simplification of Eq.~(\ref{eq:05}), we note that the density function must have a Fourier series that takes the form 
\begin{align}
f(\omega,\theta,t)=\frac{g(\omega)}{2\pi}\left(1+\sum_{n=1}^\infty\widehat{f}_n(\omega,t)e^{in\theta}+\text{c.c.}\right),\label{eq:09}
\end{align}
where c.c. denotes the complex conjugate of the preceding term. Next, using the Ott-Antonsen ansatz~\cite{Ott2008Chaos,Ott2009Chaos}, which essentially posits solutions with geometrically-decaying Fourier coefficients, i.e., $\widehat{f}_n(\omega,t)=\alpha^n(\omega,t)$, all Fourier modes remarkably reduce to a single differential equation for the function $\alpha$ given by
\begin{align}
\dot{\alpha}=-i\omega\alpha&+\frac{K_1}{2}\left(w^{(1)*}-w^{(1)}\alpha^2\right)\nonumber\\
&+\frac{K_2}{2}\left(w^{(2)*}w^{(1)}-w^{(2)}w^{(1)*}\alpha^2\right)\nonumber\\
&+\frac{K_3}{2}\left((w^{(1)*})^2w^{(1)}-(w^{(1)})^2w^{(1)*}\alpha^2\right)\label{eq:10}
\end{align}
To connect the dynamics of $\alpha$ to the order parameter, we consider the case of Lorentzian-distributed frequencies, letting
\begin{align}
g(\omega)=\frac{\Delta}{\pi[\Delta^2+(\omega-\omega_0)^2]},\label{eq:11}
\end{align}
where $\Delta$ and $\omega_0$ give the spread and mean of the natural frequencies. Specifically, for the proposed density function $f$, Eq.~(\ref{eq:06}) may first be integrated in the $\theta$ direction, yielding
\begin{align}
z^*(t)=\int\alpha(\omega,t)g(\omega d\omega.\label{eq:12}
\end{align}
Equation~(\ref{eq:12}) may further be evaluated using the Cauchy residue theorem, taking advantage of the simple pole of the frequency distribution $g(\omega)$ at $\omega=\omega_0-i\Delta$, resulting in
\begin{align}
z^*(t)=\alpha(\omega_0-i\Delta,t).\label{eq:13}
\end{align}
Thus, by evaluating Eq.~(\ref{eq:10}) at $\omega=\omega_0-i\Delta$ and taking a complex conjugate, we obtain the following differential equation for $z$:
\begin{align}
\dot{z}=-\Delta z+i\omega_0z&+\frac{K_1}{2}\left(w^{(1)}-w^{(1)*}z^2\right)\nonumber\\
&+\frac{K_2}{2}\left(w^{(2)}w^{(1)*}-w^{(2)*}w^{(1)}z^2\right)\nonumber\\
&+\frac{K_3}{2}\left((w^{(1)})^2w^{(1)*}-(w^{(1)*})^2w^{(1)}z^2\right)\label{eq:14}
\end{align}

To close the dynamics we now seek differential equations for $w^{(1)}$ and $w^{(2)}$. We begin by noting that Eqs.~(\ref{eq:07}) and (\ref{eq:08}) may be rewritten
\begin{align}
w^{(1)}(t)&=\int z(t-\tau)h(\tau)d\tau,~\text{and}\label{eq:15}\\
w^{(2)}(t)&=\int z_2(t-\tau)h(\tau)d\tau,\label{eq:16}
\end{align}
where $z_2(t)=\iint f(\omega,\theta,t)e^{2i\theta(t)}d\theta d\omega$ is the Daido order parameter~\cite{}, which following the dimensionality reduction above is simply given by $z_2(t)=z^2(t)$.
By considering the special case of exponentially-distributed time delays, namely letting
\begin{align}
h(\tau)=\left\{\begin{array}{rl}\frac{1}{T}e^{-\tau/T} &\text{if }\tau\ge0\\0&\text{if }\tau<0,\end{array}\right.\label{eq:17}
\end{align}
so that the characteristic time delay between oscillators is given by $T$, Eqs.~(\ref{eq:15}) and (\ref{eq:16}) may be treated with the Laplace transform to obtain the following differential equations:
\begin{align}
T\dot{w}^{(1)}&=z-w^{(1)},~\text{and}\label{eq:18}\\
T\dot{w}^{(2)}&=z^2-w^{(2)}.\label{eq:19}
\end{align}
Thus, Eqs.~(\ref{eq:14}), (\ref{eq:18}), and (\ref{eq:19}) constitute a closed system for the dynamics of the instantaneous and time-delayed order parameters.

\section{Steady-State Bifurcation Analysis}\label{sec3}

To proceed with our analysis of the low dimensional dynamics given by Eqs.~(\ref{eq:14}), (\ref{eq:18}), and (\ref{eq:19}), we begin by rewriting the dynamics in polar coordinates, yielding
\begin{align}
\dot{r}&=-\Delta r + \frac{1-r^2}{2}\rho^{(1)}\left[K_1\cos(\phi^{(1)}-\psi)\right.\nonumber\\
&~~~~~~~~~~~~~~~~~~~~~~~~~~\left.+K_2\rho^{(2)}\cos(\phi^{(2)}-\phi^{(1)}-\psi)\right.\nonumber\\
&~~~~~~~~~~~~~~~~~~~~~~~~~~\left.+K_3\rho^{(1)2}\cos(\phi^{(1)}-\psi)\right],\label{eq:20}\\
\dot{\psi}&=\omega_0+\frac{1+r^2}{2r}\rho^{(1)}\left[K_1\sin(\phi^{(1)}-\psi)\right.\nonumber\\
&~~~~~~~~~~~~~~~~~~~~~~~\left.+K_2\rho^{(2)}\sin(\phi^{(2)}-\phi^{(1)}-\psi)\right.\nonumber\\
&~~~~~~~~~~~~~~~~~~~~~~~\left.+K_3\rho^{(1)2}\sin(\phi^{(1)}-\psi)\right],\label{eq:21}\\
&T\dot{\rho}^{(1)}=r\cos(\psi-\phi^{(1)})-\rho^{(1)},\label{eq:22}\\
&T\dot{\phi}^{(1)}=\frac{r}{\rho^{(1)}}\sin(\psi-\phi^{(1)}),\label{eq:23}\\
&T\dot{\rho}^{(2)}=r^2\cos(2\psi-\phi^{(2)})-\rho^{(2)},\label{eq:24}\\
&T\dot{\phi}^{(2)}=\frac{r^2}{\rho^{(2)}}\sin(2\psi-\phi^{(2)}).\label{eq:25}
\end{align}
We seek steady-state solutions where the global order parameters reach a fixed amplitude, $\dot{r}=\dot{\rho}^{(1)}=\dot{\rho}^{(2)}=0$, with a phase that processes at a constant rate, $\dot{\psi}=\dot{\phi}^{(1)}=\Omega$ and $\dot{\phi}^{(2)}=2\Omega$. (Note that $\phi^{(2)}$ processes at twice the velocity as $\psi$ and $\phi^{(1)}$ since $w^{(2)}$ chases $z^2$, which processes with twice the angular velocity of $z$.) Applying this to the time-delayed order parameters equations~(\ref{eq:22}) and (\ref{eq:23}), we get
\begin{align}
\frac{\rho^{(1)}}{r}&=\cos(\psi-\phi^{(1)}),\label{eq:26}\\
\frac{\rho^{(1)}}{r}T\Omega&=\sin(\psi-\phi^{(1)}),\label{eq:27}
\end{align}
and after using the trigonometric identity $\cos^2x+\sin^2x=1$ we obtain
\begin{align}
\rho^{(1)}=\frac{r}{\sqrt{1+T^2\Omega^2}}.\label{eq:28}
\end{align}
A similar treatment of Eqs.~(\ref{eq:24}) and (\ref{eq:25}) yields
\begin{align}
\frac{\rho^{(2)}}{r^2}&=\cos(2\psi-\phi^{(2)}),\label{eq:29}\\
\frac{\rho^{(2)}}{r^2}T\Omega&=\sin(2\psi-\phi^{(2)}),\label{eq:30}
\end{align}
and subsequently,
\begin{align}
\rho^{(2)}=\frac{r^2}{\sqrt{1+4T^2\Omega^2}}.\label{eq:31}
\end{align}

Before returning to Eqs.~(\ref{eq:20}) and (\ref{eq:21}) it will also be convenient to eliminate the trigonometric quantities in those equations. The terms $\cos(\phi^{(1)}-\psi)$ and $\sin(\phi^{(1)}-\psi)$ may be eliminated simply using Eqs.~(\ref{eq:26})--(\ref{eq:28}), yielding
\begin{align}
\cos(\phi^{(1)}-\psi)&=\frac{1}{\sqrt{1+T^2\Omega^2}},\label{eq:32}\\
\sin(\phi^{(1)}-\psi)&=-\frac{T\Omega}{\sqrt{1+T^2\Omega^2}}.\label{eq:33}
\end{align}
The terms $\cos(\phi^{(2)}-\phi^{(1)}-\psi)$ and $\sin(\phi^{(2)}-\phi^{(1)}-\psi)$, on the other hand, require some more care and the use of the trigonometric identities $\cos(x+y)=\cos(x)\cos(y)-\sin(x)\sin(y)$ and $\sin(x+y)=\sin(x)\cos(y)+\cos(x)\sin(y)$. Rewriting the argument $\phi^{(2)}-\phi^{(1)}-\psi=(\psi-\phi^{(1)})+(\phi^{(2)}-2\psi)$ and using Eqs.~(\ref{eq:26})--(\ref{eq:33}) yields
\begin{align}
\cos(\phi^{(2)}-\phi^{(1)}-\psi)&=\frac{1+2T^2\Omega^2}{\sqrt{1+T^2\Omega^2}\sqrt{1+4T^2\Omega^2}},\label{eq:34}\\
\sin(\phi^{(2)}-\phi^{(1)}-\psi)&=-\frac{T\Omega}{\sqrt{1+T^2\Omega^2}\sqrt{1+4T^2\Omega^2}}.\label{eq:35}
\end{align}
We note that Eqs.~(\ref{eq:28}), (\ref{eq:31})--(\ref{eq:35}) may also be derived using Eqs.~(\ref{eq:15}) and (\ref{eq:16}). We present this alternative derivation In Appendix~\ref{appA}.

We now have the ingredients necessary for returning to Eqs.~(\ref{eq:20}) and (\ref{eq:21}). Specifically, using Eqs.~(\ref{eq:28}), (\ref{eq:31})--(\ref{eq:35}), seeking the stationary state in Eqs.~(\ref{eq:20}) and (\ref{eq:21}) yields
\begin{widetext}
\begin{align}
\Delta r &= \frac{r(1-r^2)}{2(1+T^2\Omega^2)}\left[K_1+K_2\frac{r^2(1+2T^2\Omega^2)}{1+4T^2\Omega^2}+K_3\frac{r^2}{1+T^2\Omega^2}\right],\label{eq:36}\\
\Omega &= \omega_0 - \frac{(1+r^2)T\Omega}{2(1+T^2\Omega^2)}\left[K_1+K_2\frac{r^2}{1+4T^2\Omega^2}+K_3\frac{r^2}{1+T^2\Omega^2}\right].\label{eq:37}
\end{align}
\end{widetext}
Eqs.~(\ref{eq:36}) and (\ref{eq:37}) characterize the degree of synchronization via the amplitude $r$ of the order parameter and the angular velocity $\Omega$ of the synchronized state depending on the coupling strengths $K_1$, $K_2$, and $K_3$, the characteristic time delay $T$, and the width of the frequency distribution $\Delta$.

\begin{figure*}[t]
\centering
\epsfig{file =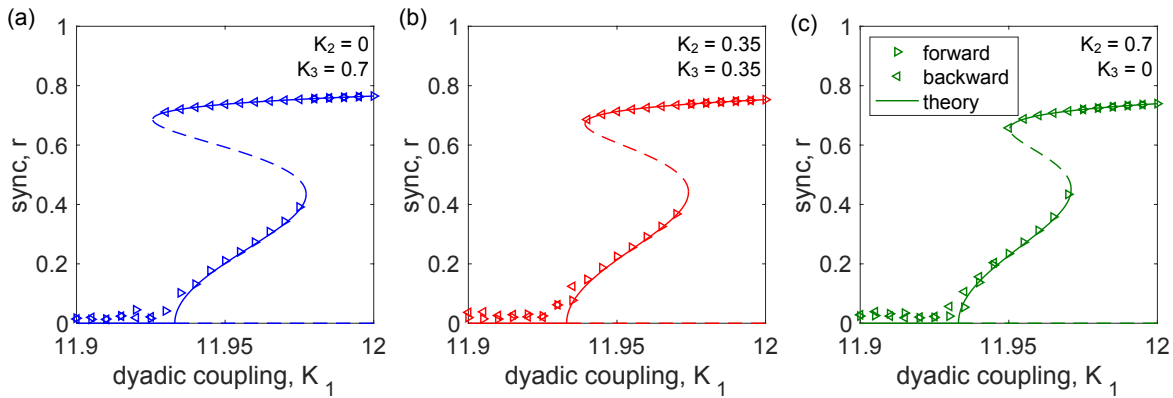, clip =,width=0.90\linewidth }
\caption{{\it Tiered synchronization.} Synchronization profiles plotting the amplitude $r$ of the order parameter vs dyadic coupling $K_1$ for three different combinations of higher order coupling: (a) $K_2=0$ and $K_3=0.7$, (b) $K_2=K_3=0.35$, and (c) $K_2=0.7$ and $K_3=0$. Results from forward and backward simulations, obtained by first adiabatically increasing then decreasing $K_1$ are plotted in forward and backward triangles, respectively, and analytical predictions obtained from solving Eqs.~(\ref{eq:36}) and (\ref{eq:37}) are plotted in solid and dashed curves, indicating stability and instability. Other parameters are $T=0.4$, $\omega_0=7.8$, and $\Delta = 1$.}\label{fig1}
\end{figure*}

To verify the dimensionality reduction and steady-state bifurcation analysis we now compare results from simulation to the solutions predicted by Eqs.~(\ref{eq:36}) and (\ref{eq:37}). To overcome the numerical complexity of incorporating explicit time delays in simulations, we consider the dynamics rewritten as Eq.~(\ref{eq:05}) with the time-delayed order parameter dynamics given by Eqs.~(\ref{eq:18}) and (\ref{eq:19}). In particular, we consider the system dynamics as dyadic coupling $K_1$ is adiabatically increased then decreased for three combinations of higher-order coupling, $K_2=0$ and $K_3=0.7$, $K_2=K_3=0.35$, and $K_2=0.7$ and $K_3=0$, and plot the simulation results using forward and backward triangles, respectively, in Figs.~\ref{fig1}(a), (b), and (c). Simulations use $N=10^4$ oscillators and at each different value of $K_1$ run through a transient of $2\times10^5$ time steps with $\Delta t=4\times10^{-3}$, then average $r$ over $2\times10^5$ time steps. We then plot the analytical predictions given by Eqs.~(\ref{eq:36}) and (\ref{eq:37}), solved numerically, using solid and dashed curves, indicating stability and instability, respectively. Other parameters are fixed at $T=0.4$, $\omega_0=7.8$, and $\Delta = 1$. We note here that the small values of $K_2$ and $K_3$ compared to $K_1$ are in line with phase reduction analyses that tend to characterize higher-order coupling as small in comparison to dyadic coupling.

The results plotted in Fig.~\ref{fig1} demonstrate a rich set of dynamics that come from the combination of time delays and higher-order interactions. In particular, for each choice of higher-order coupling used, the dynamics admit ranges of bistability. However, unlike the typical scenarios so far observed in coupled oscillator systems with time-delayed interactions and in coupled oscillator systems with higher-order interactions where bistability occurs between the incoherent state and a strongly synchronized state, here there are regions of bistability between a weakly synchronized state and a strongly synchronized state. I.e., rather than bistability occurring directly from a subcriticality and a saddle node, here supercriticality is maintained at the onset of synchronization and bistability emerges from the formation of a pair of saddle-node bifurcations. Thus, a tiered synchronization profile emerges, where a stable, weakly synchronized state, characterized by relatively small $r$, exists beyond the supercritical Hopf bifurcation at the onset of synchronization, and a stable, strongly synchronization state, characterized by larger values of $r$, exists after the curve folds over onto itself twice through a pair of saddle node bifurcations. Moreover, these two stable synchronized branches are connected via an unstable branch. 

\begin{figure}[b]
\centering
\epsfig{file =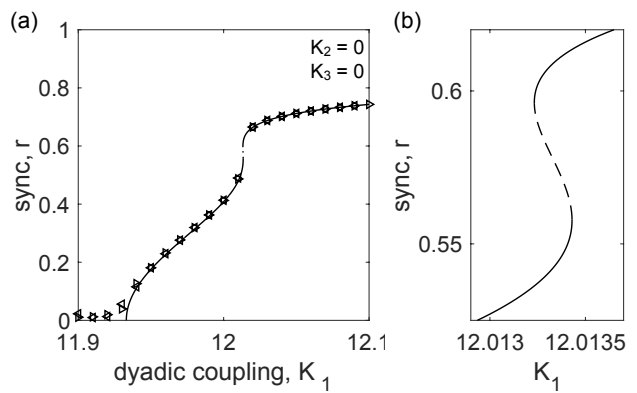, clip =,width=1.0\linewidth }
\caption{{\it Dyadic coupling only.} (a) Similar to the panels in Fig.~\ref{fig1}, the synchronization profile plotting the amplitude $r$ of the order parameter vs dyadic coupling $K_1$, but using only dyadic coupling, i.e., $K_2=K_3=0$. Results from forward and backward simulations and analytical predictions are plotted similarly. (b) A zoomed-in view of the small folded region of bistability. Other parameters are $T=0.4$, $\omega_0=7.8$, and $\Delta = 1$.}\label{fig2}
\end{figure}

For comparison, we also plot in Fig.~\ref{fig2}(a) similar results and parameters, but for the case of dyadic coupling only, i.e., with $K_2=K_3=0$. We point out that the lack of higher order interactions lead to a far less pronounced region of bistability. However, bistability does in fact exist for this choice of parameters, albeit in a very thin region of the coupling strength $K_1$. In Fig.~\ref{fig2}(b) we present a zoomed-in view of this region. The presence of this bistability deserves a few remarks. First, to our knowledge, bistability of this nature (i.e., owing to a pair of saddle nodes after a supercritical pitchfork o Hopf bifuraction) has not been observed in systems with only interaction delays (i.e., without higher-order interactions). Second, these bistability regions appear to be so thin that they are virtually unobservable in direct simulations of oscillator systems, even for large enough systems where finite-fluctuations have been all but eliminated. Note that the simulation results in Fig.~\ref{fig2}(a) show no trace of bistability--for this we need the analytically predicted curve. Taking these two points together, it appears that the presence of higher order interactions remains an important ingredient for bistability of this nature in oscillator systems. In the next two sections we, respectively, analyze the incoherent state and consider a special case that allows us to sketch the bifurcation diagram of the system.

\section{Incoherent State}\label{sec4}

Before analyzing the nonlinear effects that are present in the dynamics, we first focus our attention on the incoherent state described by $z=w^{(1)}=w^{(2)}=0$. Note that our steady state Eq.~(\ref{eq:36}) implies that the incoherent state is always a solution, and so the stability properties of the incoherent state determines the onset of synchronization. By eliminating the incoherent state from Eq.~(\ref{eq:36}) and letting $r\to0^+$ in Eqs.~(\ref{eq:36}) and (\ref{eq:37}) we obtain a simplified set of equations describing this critical point corresponding to the collision between the synchronized an incoherent branches in Fig.~\ref{fig1}, given by
\begin{align}
\Delta &=\frac{K_1}{2(1+T^2\Omega^2)},\label{eq:38}\\
\Omega&=\omega_0-\frac{K_1T\Omega}{2(1+T^2\Omega^2)}.\label{eq:39}
\end{align}
Eqs.~(\ref{eq:38}) and (\ref{eq:39}) may be combined to find
\begin{align}
\Omega=\frac{\omega_0}{1+T\Delta},\label{eq:40}
\end{align}
which may be inserted back in to Eq.~(\ref{eq:38}) and solved for $K_1$ to yield the critical dyadic coupling strength
\begin{align}
K_1^c=2\Delta+\frac{2\Delta T^2\omega_0^2}{(1+T\Delta)^2}.\label{eq:41}
\end{align}
We note that from Eq.~(\ref{eq:41}) we see that this critical coupling strength depends monotonically on both the mean natural frequency $\omega_0$ and the characteristic time delay $T$, but not the width of the frequency distribution, $\Delta$. I.e. increasing either $\omega_0$ or $T$ causes $K_1^c$ to increase, but increasing $\Delta$ may increase or decrease $K_1^c$.

Alternatively, it is easy to check that this critical coupling strength corresponds exactly to the first crossing of the the eigenvalues of the complex-valued Jacobian of Eqs.~(\ref{eq:14}), (\ref{eq:18}), and (\ref{eq:19}) for the incoherent state, given by
\begin{align}
DF=\begin{bmatrix}-\Delta +i\omega_0 & \frac{K_1}{2} & 0\\ \frac{1}{T}&-\frac{1}{T}& 0 \\ 0&0&-\frac{1}{T}\end{bmatrix}.\label{eq:42}
\end{align}
Specifically, the incoherent state is asymptotically stable (with all eigenvalues located in the left-half complex plane) for $K_1<K_1^c$, after which stability is lost at $K_1=K_1^c$. Given the rotating nature of solutions (described by the angular velocity $\Omega$) this is a Hopf bifurcation that may be either supercritical or subcritical, depending on the nature of the steady-state solutions (In Fig.~\ref{fig1} it is supercritical for all thee cases). However, in the appropriate rotating reference frame it may viewed as a pitchfork bifurcation. In Fig.~\ref{fig1} we delineate the stable and unstable portions of the incoherent branch in solid and dashed curves along $r=0$. Importantly, we note that the onset of synchronization is unaffected by the triadic and tetradic coupling strengths, $K_2$ and $K_3$, implying that in terms of the macroscopic dynamics the higher order interactions offer only nonlinear effects.

\section{Synchronized States without Triadic Coupling}\label{sec5}

\begin{figure}[t]
\centering
\epsfig{file =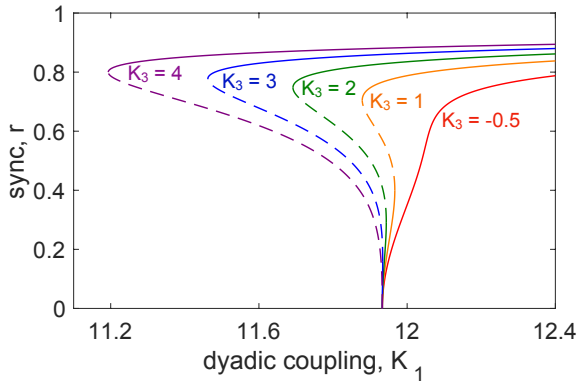, clip =,width=0.95\linewidth }
\caption{{\it Synchronized states without triadic coupling.} Synchronization profiles $r$ vs $K_1$ in the presence of only dyadic and tetradic coupling, i.e., $K_2=0$, given by Eq.~(\ref{eq:41}) for $K_3=-0.5$, $1$, $2$, $3$, and $4$ (red to purple, right to left). Other parameters are $T=0.4$, $\omega_0=7.8$, and $\Delta = 1$.}\label{fig3}
\end{figure}

We now turn our attention to the special case where triadic coupling is eliminated, i.e., we set $K_2=0$, leaving only dyadic and tetradic coupling. As we will see, this special case allows for a modest simplification of the general case, thereby allowing for further analytical results and a fuller picture of the bifurcation diagram. We begin by noting that when $K_2$ is set to zero, Eqs.~(\ref{eq:36}) and (\ref{eq:37}) reduce to
\begin{align}
\Delta r &= \frac{r(1-r^2)}{2(1+T^2\Omega^2)}\left(K_1+K_3\frac{r^2}{1+T^2\Omega^2}\right),\label{eq:43}\\
\Omega &= \omega_0 - \frac{(1+r^2)T\Omega}{2(1+T^2\Omega^2)}\left(K_1+K_3\frac{r^2}{1+T^2\Omega^2}\right),\label{eq:44}
\end{align}
which may be combined to solve for $\Omega$, yielding
\begin{align}
\Omega = \frac{\omega_0}{1+\frac{1+r^2}{1-r^2}\Delta T}.\label{eq:45}
\end{align}
Inserting Eq.~(\ref{eq:45}) into Eq.~(\ref{eq:43}) and solving for $K_1$ then yields
\begin{align}
K_1&=\frac{2\Delta+\frac{2\Delta T^2\omega_0^2}{\left(1+\frac{1+r^2}{1-r^2}\Delta T\right)^2}}{1-r^2}-\frac{K_3r^2}{1+\frac{T^2\omega_0^2}{\left(1+\frac{1+r^2}{1-r^2}\Delta T\right)^2}}.\label{eq:46}
\end{align}
While Eq.~(\ref{eq:46}) describes $K_1$ as a function of $r$ (rather than vice-versa) it turns out to be extremely useful for exploring the system dynamics. First, we use it to plot multiple synchronization profiles in Fig.~\ref{fig3} for $K_3=-0.5$, $1$, $2$, $3$, and $4$ (red to purple, right to left). Other parameters are the same as those used in Fig.~\ref{fig1}, except for $K_2=0$. In fact, these example choices of $K_3$ sweep out a collection of curves that illustrate a rich range of system dynamics. In particular, starting at $K_3=-0.5$ (right-most curve), the system undergoes a single supercritical Hopf bifurcation delineating the incoherent a synchronized states. As $K_3$ is then increased the dynamics qualitatively change, as the top portion of the curve develops a fold over onto itself, but then folds back. This folding corresponds to a pair of saddle node bifurcation and separates a single stable synchronized branch into two stable synchronized branches, corresponding to weak and strong synchronization. These branches are connected by another branch that is unstable. This is illustrated by the $K_3=1$, $2$, and $3$ curves. Finally, for even larger $K_3$ the bottom saddle-node disappears as the weakly synchronized branch vanishes as the onset of synchronization becomes a subcritical Hopf bifurcation, as illustrated by the $K_3=4$ curve.

\begin{figure}[t]
\centering
\epsfig{file =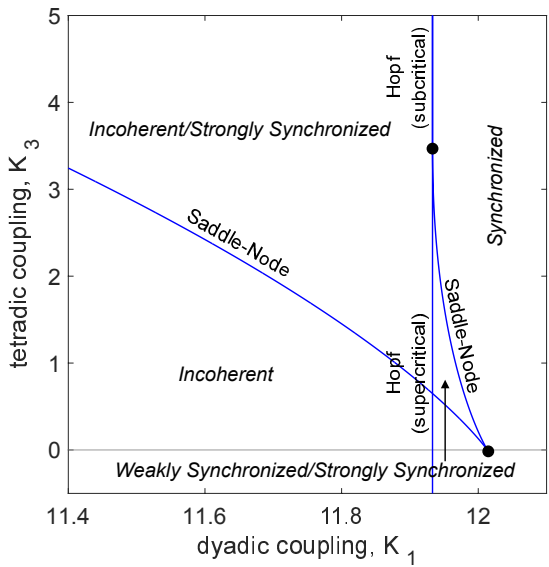, clip =,width=0.95\linewidth }
\caption{{\it Bifurcation diagram.} Bifurcation diagram of the system for dyadic and tetradic coupling $K_1$ and $K_2$ in the absence of triadic coupling, i.e., $K_2=0$. Bifurcation curves are labeled accordingly and the stable states in each region of state space (incoherent and synchronized) are indicated in italic text. Two codimension-two points are shown as black circles. Other parameters are $T=0.4$, $\omega_0=7.8$, and $\Delta = 1$.}\label{fig4}
\end{figure}

Moreover, while Eq.~(\ref{eq:46}) does not characterize $r$ as a function of $K_1$, but rather vice-versa, for purposes of sketching the bifurcation diagram of the system this format is quite convenient, since the saddle node bifurcations described in Figs.~\ref{fig1} and \ref{fig3} may be found and described using the derivative $\frac{\partial K_1}{\partial r}$. Treating Eq.~(\ref{eq:41}) analytically remains difficult, but the appropriate bifurcation conditions may be easily identified numerically by searching for local minima and maxima of $K_1$ as a function of $r$. In Fig.~\ref{fig4} we plot the bifurcation diagram over dyadic coupling $K_1$ and tetradic coupling $K_3$. Bifurcation curves indicate Hopf and saddle-node bifurcations, labeled accordingly. In fact, the pair of saddle-node bifurcations are born at a codimension-two point, indicated by the lower black circle. The saddle node curves split as $K_3$ increases, and eventually the higher saddle-node curve collides with the pitchfork bifurcation at another codimension-two point, above and below which the Hopf is subcritical and supercritical.

In addition to providing a representatively full picture of the macroscopic dynamics, the bifurcation diagram in Fig.~\ref{fig4} demonstrates the point made previously where bistability owing to a supercritical Hopf bifurcation followed by a pair of saddle nodes can be achieved in the absence of higher order interactions, but in fact is so subtle that it is difficult to observe in simulations and occurs for such a thin region of $K_1$. In particular, the codimension-two point at the formation of the two saddle nodes in Fig.~\ref{fig4} lies just under the $K_3=0$ line, indicating that at $K_3=0$, i.e., in the absence of higher order interactions, the pair of saddle nodes exists (for the parameters chosen). Again, we emphasize that to our knowledge this kind of transition has not been observed in previous work, likely due to the thin parameter range of bistability (see also Fig.~\ref{fig1}).

\section{Discussion}\label{sec6}

In this paper we have studied the synchronization dynamics of populations of coupled phase oscillators with both interaction delays and higher order interactions. After employing the dimensionality reduction of Ott and Antonsen~\cite{Ott2008Chaos,Ott2009Chaos} we presented an analysis of the steady-state solutions. We showed that the combination of interactions delays and higher-order interactions promote a bistability that, unlike what has been observed previously when either interactions delays or higher-order interactions are present, leaves the onset of synchronization supercritical and creates two stable synchronized states via a pair of saddle nodes. 

We also highlight two other pieces of interest that emerge from our analysis. First, as is evident from the contribution of triadic and tetradic coupling strengths in low dimensional equations, the presence of higher-order interactions offer only nonlinear effects to the macroscopic system dynamics. This is further emphasized by the fact that the onset of synchronization does not depend on the triadic or tetradic coupling strengths. Second, while mechanism for bistability that gives rise to tiered synchronization discussed above is promoted by the combination of interactions delays and higher order interactions, we find that the double saddle node can in fact be observed with only interactions delays. It appears that the regions of parameter space that give rise to such transitions are simply very small and have not been previously observed.

Lastly, we have considered here the case of heterogeneous oscillators and heterogeneous time delays, where natural frequencies and time delays are distributed via a Lorentzian distribution and an exponential distribution, respectively. While these choices were made in order to facilitate analytical treatment of the system, specifically aiding in the derivation of the low dimensional dynamics, it remains to be seen whether different choices of natural frequency distributions and/or interaction delays qualitatively change the macroscopic dynamics, a task left for future work.

\acknowledgments
PSS acknowledges support from NSF grant MCB-2126177. CX acknowledges National Natural Science Foundation of China grant No. 11905068 and the Scientific Research Funds of Huaqiao University grant No. ZQN-810.

\appendix

\section{Alternative derivation of Eqs.~(\ref{eq:28}), (\ref{eq:31})--(\ref{eq:35})}\label{appA}

We begin our alternative derivation with Eq.~(\ref{eq:15}). Using the polar form for $z$ and $w^{(1)}$, choosing $\psi=\Omega t$, and denoting $\varphi^{(1)}=\psi-\phi^{(1)}$ we have that
\begin{align}
\rho^{(1)}e^{i\phi^{(1)}}&=\int r(t-\tau)e^{i\psi(t-\tau)}h(\tau)d\tau\label{eq:appA}\\
&=\int re^{i\Omega(t-\tau)}h(\tau)d\tau\label{eq:appB}\\
&=re^{i\Omega t}\int e^{-i\Omega\tau}h(\tau)d\tau\label{eq:appC}\\
&=\frac{re^{i\Omega t}}{1+iT\Omega}\label{eq:appD}\\
&=\frac{re^{i(\Omega t-\varphi^{(1)})}}{\sqrt{1+T^2\Omega^2}}\label{eq:appE}
\end{align}
Taking an absolute value then recovers Eq.~(\ref{eq:28}) in the main text. Moreover, taking real an imaginary parts, we recover Eqs.~(\ref{eq:32}) and (\ref{eq:33}).

On the other hand, if we begin with Eq.~(\ref{eq:16}) we may similarly write
\begin{align}
\rho^{(2)}e^{i\phi^{(2)}}&=\int r^2(t-\tau)e^{i2\psi(t-\tau)}h(\tau)d\tau\label{eq:appF}\\
&=\int r^2e^{2i\Omega(t-\tau)}h(\tau)d\tau\label{eq:appG}\\
&=r^2 e^{2i\Omega t}\int e^{-2i\Omega\tau}h(\tau)d\tau\label{eq:appH}\\
&=\frac{r^2 e^{2i\Omega t}}{1+2iT\Omega}\label{eq:appI}\\
&=\frac{r^2 e^{i(2\Omega t-\varphi^{(2)})}}{\sqrt{1+4T^2\Omega^2}}\label{eq:appJ}
\end{align}
Taking an absolute value then recovers Eq.~(\ref{eq:31}) in the main text. Moreover, taking real an imaginary parts, we obtain
\begin{align}
\cos(\phi^{(2)}-2\psi)&=\frac{1}{\sqrt{1+4T^2\Omega^2}},\label{eq:appK}\\
\sin(\phi^{(2)}-2\psi)&=\frac{-2T\Omega}{\sqrt{1+4T^2\Omega^2}},\label{eq:appL}
\end{align}
which can be used along with Eqs.~(\ref{eq:32}) and (\ref{eq:33}) to yield Eqs.~(\ref{eq:34}) and (\ref{eq:35}) in the main text.

\bibliographystyle{plain}

\end{document}